\documentclass[a4paper,12pt,twoside]{article}

\usepackage[dvips]{graphicx}
\usepackage{geometry}
\usepackage[centertags]{amsmath}
\usepackage{amsfonts}
\usepackage{amssymb}
\usepackage{latexsym}
\usepackage{mathrsfs}
\usepackage{fancyhdr}
\usepackage[sort&compress,comma,square]{natbib}

\RequirePackage[pdfstartview=FitH]{hyperref}

\hypersetup{%
  pdftitle={Action Principle and Modification of the
  Faddeev-Popov Factor in Gauge Theories},%
  pdfsubject={Gauge Field Theories},%
  pdfauthor={Kanchana Limboonsong and Edouard Berge Manoukian},%
  pdfkeywords={Schwinger (Quantum) Action Principle,
  Gauge Theories, Faddeev-Popov Factor and Quantization Rules},%
}

\newcommand{\email}[1]{\href{mailto:#1}{\texttt{#1}}}

\newcommand{\braket}[2]{\ensuremath{\left\langle{#1}\!%
\mathrel{\left|{\vphantom{{#1} {#2}}}\right.%
\kern-\nulldelimiterspace}\!{#2}\right\rangle}}

\newcommand{\ketbra}[2]{\ensuremath{\left|{#1}\!%
\mathrel{\left\rangle\vphantom{{#1} {#2}}\right\langle%
\kern-\nulldelimiterspace}\!{#2}\right|}}

\newcommand{\BK}[3]{\ensuremath{\left\langle{#1}\!%
\mathrel{\left|\vphantom{{#1}}{#2}\vphantom{{#3}}\right|%
\kern-\nulldelimiterspace}\!{#3}\right\rangle}}

\newcommand{\commute}[2]{\ensuremath{\left[{#1}\!%
\mathrel{\vphantom{{#1}},\vphantom{{#2}}%
\kern-\nulldelimiterspace}\!{#2}\right]}}

\newcommand{\Vacuum}{\ensuremath{\braket{0_{+}}{0_{-}}}}

\newcommand{\uI}{\ensuremath{\mathrm{i}}}
\newcommand{\uE}{\ensuremath{\mathrm{e}}}
\newcommand{\uD}{\ensuremath{\mathrm{d}}}

\newcommand{\Tr}{\ensuremath{\operatorname{Tr}}}

\renewcommand{\Vec}[1]{\ensuremath{\boldsymbol{\vec{\mathrm{#1}}}}}

\def\Beginboxit{\par\vbox\bgroup\hrule\hbox\bgroup%
                \vrule \kern1.2pt \vbox\bgroup\kern1.2pt}
\def\Endboxit{\kern1.2pt\egroup\kern1.2pt\vrule\egroup%
              \hrule\egroup}{}
\newenvironment{boxit}{\Beginboxit}{\Endboxit}
\newenvironment{boxit*}{\Beginboxit\hbox to\hsize{}}{\Endboxit}

\newcommand{\ArXivSite}{http://arxiv.org/abs/}
\newcommand{\MyArXivNo}{0709.4603}
\newcommand{\ArXivNo}[1]{\href{\ArXivSite#1}{#1}}

\newcommand{\XXXSize}{{\fontsize{12}{12}\selectfont\fbox{%
\textbf{arXiv:\ArXivNo{\MyArXivNo}v1 [hep-th]}}}}
\newcommand{\XXXTitle}{\hfill\XXXSize\newline\vskip 0.4cm}

\makeatletter
\let\@afterindentfalse\@afterindenttrue
\@afterindenttrue \makeatother

\begin{document}
\pagestyle{fancy}

\allowdisplaybreaks

\sloppy

\title{\XXXTitle\textbf{Action Principle and Modification of the
Faddeev--Popov Factor in Gauge Theories}\footnote{%
Published in \textit{International Journal of Theoretical Physics}, %
Vol.~\textbf{45}, No.~10, pp.~1814--1824 (2006). \
[doi:\href{http://dx.doi.org/10.1007/s10773-006-9152-x}%
{10.1007/s10773-006-9152-x}] %
}}

\author{\textsc{Kanchana~Limboonsong} \ and \
\textsc{Edouard~B.~Manoukian}\footnote{E-mail:~\email{edouard@sut.ac.th}} \\
{\href{http://physics3.sut.ac.th/}{School of Physics}, \ %
\href{http://www.sut.ac.th/science/}{Institute of Science}} \\
{\href{http://www.sut.ac.th/}{Suranaree University of Technology}} \\
{Nakhon~Ratchasima, 30000, Thailand}}

\date{}

\maketitle

\begin{boxit}
\begin{abstract}
The quantum action (dynamical) principle is exploited to
investigate the nature and origin of the Faddeev--Popov (FP)
factor in gauge theories without recourse to path integrals.
Gauge invariant as well as gauge non-invariant interactions are
considered to show that the FP factor needs to be modified in more
general cases and expressions for these modifications are derived.
In particular we show that a gauge invariant theory does not
necessarily imply the familiar FP factor for
proper quantization. 
\begin{flushleft}
\noindent\textbf{Key Words:} action principle, gauge
theories, Faddeev--Popov factor and quantization rules. \\
\noindent\textbf{PACS Numbers:} \texttt{11.15.-q, 12.10.-g, 12.15.-y,
12.38.-t}
\end{flushleft}
\end{abstract}
\end{boxit}

\section{Introduction}\label{Section1}

In earlier communications~\cite{Manoukian_1986,Manoukian_1987,
Manoukian_2005}, we have seen that the quantum action (dynamical)
principle~\cite{Schwinger_1951a,Schwinger_1951b,Schwinger_1953a,
Schwinger_1953b,Schwinger_1954,Schwinger_1972,Schwinger_1973,
Lam_1965,Manoukian_1985} may be used to quantize gauge theories in
constructing the vacuum-to-vacuum transition amplitude and the
Faddeev--Popov (FP) factor~\cite{Faddeev_1967}, encountered in
non-abelian gauge theories (e.g., \cite{Abers_1973,Rivers_1987,
tHooft_2000,Veltman_2000,Gross_2005,Politzer_2005,Wilczek_2005}),
may be obtained \emph{directly} from the action principle without
much effort. No appeal was made to path integrals, and there was
not even the need to go into the well-known complicated structure
of the Hamiltonian~\cite{Fradkin_1970} in non-abelian gauge
theories. For extensive references on the gauge problem in gauge
theories see Manoukian and Siranan~\cite{Manoukian_2005}.   The
latter reference traces its historical development from early
papers to most recent ones.

In the present investigation, we consider the generic non-abelian
gauge theory Lagrangian density
\begin{equation}\label{Eqn01}
  \mathscr{L}_{\mathrm{T}} = \mathscr{L}+\mathscr{L}_{\mathrm{S}}
\end{equation}
and modifications thereof, where
\begin{align}
  \mathscr{L} &= -\frac{1}{4}G^{a}_{\mu\nu}G_{a}^{\mu\nu}
  +\frac{1}{2\uI}\left[\left(\partial_{\mu}\overline{\psi}\right)
  \gamma^{\mu}\psi-\overline{\psi}\gamma^{\mu}\partial_{\mu}\psi\right]
  -m_{0}\overline{\psi}\psi
  \nonumber \\
  &\qquad{} +g_{0}\overline{\psi}\gamma_{\mu}A^{\mu}\psi
  \label{Eqn02} \\[0.5\baselineskip]
  \mathscr{L}_{\mathrm{S}} &= \overline{\eta}\psi+\overline{\psi}\eta
  +J_{a}^{\mu}A^{a}_{\mu}
  \label{Eqn03} \\[0.5\baselineskip]
  A_{\mu} &= A^{a}_{\mu}t_{a}, \qquad{}
  G_{\mu\nu} = \partial_{\mu}A_{\nu}-\partial_{\nu}A_{\mu}
  -\uI{}g_{0}\commute{A_{\mu}}{A_{\nu}}
  \label{Eqn04} \\[0.5\baselineskip]
  G_{\mu\nu} &= G^{a}_{\mu\nu}t_{a}
  \label{Eqn05} \\[0.5\baselineskip]
  G^{a}_{\mu\nu} &= \partial_{\mu}A^{a}_{\nu}-\partial_{\nu}A^{a}_{\mu}
  +g_{0}f^{abc}A^{b}_{\mu}A^{c}_{\nu}.
  \label{Eqn06}
\end{align}
The $t^{a}$ are generators of the underlying algebra, and the
$f^{abc}$, totally antisymmetric, are the structure constants
satisfying the Jacobi identity,
$\commute{t^{a}}{t^{b}}=\uI{}f^{abc}t^{c}$.   Note that $A_{\mu}$
is a matrix.   $\mathscr{L}_{\mathrm{S}}$ is the source term with
the $J^{a}_{\mu}$ classical functions, while $\eta$,
$\overline{\eta}$ are so-called anti-commuting Grassmann
variables.

The Lagrangian density $\mathscr{L}$ in (\ref{Eqn02}) is invariant
under simultaneous local gauge transformations:
\begin{align}
  \psi &\longrightarrow{} U\psi, \qquad{}
  \overline{\psi} \longrightarrow{} \overline{\psi}U^{-1},
  \label{Eqn07} \\[0.5\baselineskip]
  A_{\mu} &\longrightarrow{} UA_{\mu}U^{-1}+\frac{\uI}{g_{0}}
  U\partial_{\mu}U^{-1}
  \label{Eqn08} \\[0.5\baselineskip]
  G_{\mu\nu} &\longrightarrow{} UG_{\mu\nu}U^{-1}
  \label{Eqn09}
\end{align}
where $U=U(\theta)=\exp\left[\uI{}g_{0}\theta^{a}t^{a}\right]$,
$\theta=\theta^{a}t^{a}$, $\theta=\theta(x)$.

Upon setting
\begin{equation}\label{Eqn10}
  \nabla_{\!\!\mu} = \partial_{\mu}-\uI{}g_{0}A_{\mu}
\end{equation}
with
\begin{equation}\label{Eqn11}
  \nabla^{ab}_{\!\!\mu} = \delta^{ab}\partial_{\mu}
  +g_{0}f^{acb}A^{c}_{\mu}
\end{equation}
we have the basic commutator
\begin{equation}\label{Eqn12}
  \commute{\nabla_{\!\!\mu}}{\nabla_{\!\!\nu}} = -\uI{}g_{0}G_{\mu\nu}
\end{equation}
and the identity
\begin{equation}\label{Eqn13}
  \nabla^{ab}_{\!\!\mu}\nabla^{bc}_{\!\!\nu}G_{c}^{\mu\nu} = 0.
\end{equation}
[The latter generalizes the elementary identity
$\partial_{\mu}\partial_{\nu}F^{\mu\nu}=0$, in abelian gauge
theory, to non-abelian ones, where
$F^{\mu\nu}=\partial^{\mu}\!A^{\nu}-\partial^{\nu}\!A^{\mu}$.]

We consider gauge invariant (Sect.~\ref{Section3}) as well as
gauge non-invariant (Sect.~\ref{Section4}) modifications of the
Lagrangian density and show by a systematic use of the quantum
action principle that the familiar FP factor needs to be modified
in more general cases and explicit expressions for these
modifications are derived.   In particular, we show that a gauge
invariant theory does \emph{not} necessarily imply the familiar FP
factor for proper quantization, as may be perhaps expected (cf.
Rivers~\cite{Rivers_1987}, p.~204), and modifications thereof may
be necessary.   Before doing so, however, we use the action
principle to derive, in Sect.~\ref{Section2}, the FP factor and
investigate its origin for the classic Lagrangian density
$\mathscr{L}$, without recourse to path integrals, as an
anticipation of what to expect in more general cases.  Throughout,
we work in the celebrated Coulomb gauge $\partial_{k}A_{a}^{k}=0$,
$k=1,2,3$.

\section{Action Principle and the Origin of the FP Factor}\label{Section2}

To obtain the expression for the vacuum-to-vacuum transition
amplitude $\Vacuum$, in the presence of external sources
$J^{a}_{\mu}$, $\eta^{a}$, $\overline{\eta}^{a}$, as the generator
of all the Green functions of the theory, \emph{no} restrictions
may be set, in particular, on the external current $J^{a}_{\mu}$,
coupled to the gauge fields $A_{a}^{\mu}$, such as
$\partial^{\mu}\!J^{a}_{\mu}=0$, so that \emph{variations of the
components of $J^{a}_{\mu}$ may be carried out independently},
until the entire analysis is completed, and all functional
differentiations are carried out to generate Green functions. This
point cannot be overemphasized.   As we will see, the
\emph{generality} condition that must be adopted on the external
current $J^{a}_{\mu}$ together with the presence of
\emph{dependent} gauge field components in ($A_{a}^{\mu}$), as a
result of the structure of the Lagrangian density $\mathscr{L}$ in
(\ref{Eqn02}) and the gauge constraint, are responsible for the
\emph{origin} and the presence of the FP factor in the theory for
a proper quantization in the realm of the quantum action
principle.

We define the Green operator $D^{ab}(x,x')$ satisfying the
differential equation
\begin{equation}\label{Eqn14}
  \left[\delta^{ac}\Vec{\partial}^{2}+g_{0}f^{abc}A^{b}_{k}
  \partial_{k}\right]D^{cd}(x,x') = \delta^{4}(x,x')\delta^{ad}.
\end{equation}
Since the differential operator on the left-hand side of
$D^{cd}(x,x')$ is independent of the time derivative,
$D^{cd}(x,x')$ involves a $\delta(x^{0}-{x'}^{0})$ factor.   Using
the gauge constraint, one may, for example, eliminate $A_{a}^{3}$
in favor of $A_{a}^{1}$, $A_{a}^{2}$.   That is, we may treat the
$A_{a}^{3}$ as dependent fields.

The field equations are given by
\begin{equation}\label{Eqn15}
  \nabla^{ab}_{\!\!\mu}G_{b}^{\mu\nu} = -\left(\delta^{\nu}{}_{\sigma}
  \,\delta^{ac}-g^{\nu{}k}\partial_{k}D^{ab}\nabla^{bc}_{\!\!\sigma}
  \right)\left[J_{c}^{\sigma}+g_{0}\overline{\psi}\gamma^{\sigma}
  t_{c}\psi\right]
\end{equation}
with $\mu,\nu=0,1,2,3$, $k=1,2,3$, and
\begin{align}
  \bigg[\gamma^{\mu}\frac{\nabla_{\!\!\mu}}{\uI}+m_{0}\bigg]
  \psi &= \eta
  \label{Eqn16} \\[0.5\baselineskip]
  \overline{\psi}\bigg[\gamma^{\mu}\frac{\overset{\lower0.7em
  \hbox{$\smash[t]{\scriptstyle\leftarrow}$}}{\nabla}_{\!\!\mu}
  \!{}^{*}}{\uI}-m_{0}\bigg] &= -\overline{\eta}
  \label{Eqn17}
\end{align}
where $\nabla_{\!\!\mu}$ is defined in (\ref{Eqn10}).

The canonical conjugate variables to $A_{a}^{1}$, $A_{a}^{2}$, are
given by
\begin{equation}\label{Eqn18}
  \pi_{a}^{i} =
  G_{a}^{i0}-\partial_{3}^{-1}\partial^{i}G_{a}^{30},
  \qquad{} i=1,2.
\end{equation}
With $\pi_{a}^{0}=0$, $\pi_{a}^{3}=0$, we may rewrite
(\ref{Eqn18}) as
\begin{equation}\label{Eqn19}
  \pi_{a}^{\mu} = G_{a}^{\mu{}0}-\partial_{3}^{-1}
  g^{\mu{}k}\partial_{k}G_{a}^{30}
\end{equation}
$k=1,2,3$.   One may then readily express $G_{a}^{\mu{}0}$ as
follows:
\begin{equation}\label{Eqn20}
  G_{a}^{\mu{}0} = \pi_{a}^{\mu}-g^{\mu{}k}\partial_{k}D_{ab}
  \left[J_{b}^{0}+g_{0}\overline{\psi}\gamma^{0}t_{b}\psi
  +\nabla^{bc}_{\!\!\nu}\pi_{c}^{\nu}\right].
\end{equation}
We note that the right-hand side of (\ref{Eqn20}) is expressed in
terms of the independent fields $A_{a}^{1}$, $A_{a}^{2}$, their
canonical conjugate momenta and involves no time derivatives.
Here we recall that $A_{a}^{3}$ is expressed in terms of
$A_{a}^{1}$, $A_{a}^{2}$ with no time derivative.    Accordingly,
with the (independent) fields and their canonical conjugate
momenta kept \emph{fixed}, we obtain the following functional
derivative
\begin{equation}\label{Eqn21}
  \frac{\delta}{\delta{}J_{b}^{\nu}(x')}G_{a}^{\mu{}0}(x)
  = -g^{\mu{}k}\delta^{0}{}_{\nu}\,\partial_{k}D_{ab}(x,x')
\end{equation}
$\mu,\nu=0,1,2,3$; $k=1,2,3$.   On the other hand,
$G_{a}^{kl}=\partial^{k}A_{a}^{l}-\partial^{l}A_{a}^{k}%
+g_{0}f_{abc}A_{b}^{k}A_{c}^{l}$;
$k,l=1,2,3$, may be expressed in terms of the independent fields
$A_{a}^{1}$, $A_{a}^{2}$ and involves no time derivatives.
Accordingly with $A_{a}^{1}$, $A_{a}^{2}$ and their canonical
conjugate variables kept fixed, we also have
\begin{equation}\label{Eqn22}
  \frac{\delta}{\delta{}J_{b}^{\nu}(x')}G_{a}^{kl}(x) = 0.
\end{equation}
Similarly, with $\psi$ and $\overline{\psi}$ kept fixed, we have
the obvious functional derivative expression
\begin{equation}\label{Eqn23}
  \frac{\delta}{\delta{}J_{b}^{\nu}(x')}\left[
  \overline{\psi}(x)\gamma^{\mu}t^{a}\psi(x)\right] = 0.
\end{equation}

The action principle gives
\begin{equation}\label{Eqn24}
  \frac{\partial}{\partial{}g_{0}}\Vacuum = \uI\BK{0_{+}}{
  \int\!(\uD{}x)\,\hat{\mathscr{L}}_{\mathrm{I}}\,}{0_{-}}
\end{equation}
where
\begin{equation}\label{Eqn25}
  \hat{\mathscr{L}}_{\mathrm{I}}
  = \frac{\partial}{\partial{}g_{0}}\mathscr{L}
  = -\frac{1}{2}f^{abc}A^{b}_{\mu}A^{c}_{\nu}G_{a}^{\mu\nu}
  +\overline{\psi}\gamma^{\mu}A_{\mu}\psi.
\end{equation}
We may also write
\begin{equation}\label{Eqn26}
  f^{abc}A^{b}_{\mu}A^{c}_{\nu}G_{a}^{\mu\nu}
  = 2f^{abc}A^{b}_{k}A^{c}_{0}G_{a}^{k0}
  +f^{abc}A^{b}_{k}A^{c}_{l}G_{a}^{kl}
\end{equation}
and set $(-\uI)\delta/\delta{}J_{a}^{\mu}={A'}^{a}_{\!\!\mu}$,
$(-\uI)\delta/\delta\overline{\eta}=\psi'$,
$(-\uI)\delta/\delta\eta=\overline{\psi}'$.    [Here we note that
${G'}^{a}_{\!\!\mu\nu}$ on the right-hand side of (5.30) of
Manoukian~\cite{Manoukian_1986} should be replaced by
${F'}^{a}_{\!\!\mu\nu}=\partial_{\mu}{A'}^{a}_{\!\!\nu}
-\partial_{\nu}{A'}^{a}_{\!\!\mu}$.]

Now we use the rule of functional differentiations
(Lam~\cite{Lam_1965};
Manoukian~\cite{Manoukian_1985,Manoukian_1986,
Manoukian_1987,Manoukian_2006}) that for an operator
$\mathcal{O}(x)$
\begin{align}
  (-\uI)\frac{\delta}{\delta{}J_{a}^{\mu}(x')}
  \BK{0_{+}}{\mathcal{O}(x)}{0_{-}}
  &= \BK{0_{+}}{\big(A^{a}_{\mu}(x')
  \mathcal{O}(x)\big)_{+}}{0_{-}}
  \nonumber \\
  &\qquad{}
  -\uI\BK{0_{+}}{\frac{\delta}{\delta{}
  J_{a}^{\mu}(x')}\mathcal{O}(x)}{0_{-}}
  \label{Eqn27}
\end{align}
where $(\ldots)_{+}$ denotes the time-ordered product, and the
functional derivative of $\mathcal{O}(x)$ in the second term on
the right-hand of (\ref{Eqn27}) is taken as in
(\ref{Eqn21})--(\ref{Eqn23}) with the (independent) fields and
their canonical conjugate momenta kept fixed.    Here we recall
that $A_{a}^{3}$ may be expressed in terms of $A_{a}^{1}$,
$A_{a}^{2}$ and involves no time derivatives.

From (\ref{Eqn24})--(\ref{Eqn27}), together with
(\ref{Eqn21})--(\ref{Eqn23}), we obtain
\begin{equation}\label{Eqn28}
  \frac{\partial}{\partial{}g_{0}}\Vacuum = \int\!(\uD{}x)\left[
  \uI{\hat{\mathscr{L}}}'_{\mathrm{I}}(x)-f^{bca}
  {A'}^{b}_{\!\!k}\partial^{k}{D'}^{ac}(x,x)\right]\Vacuum.
\end{equation}
Using a matrix notation
\begin{equation}\label{Eqn29}
  D^{ab}(x,x') = \left[\BK{x}{\left(
  \frac{1}{\Vec{\partial}^{2}-\uI{}g_{0}A_{k}\partial_{k}}
  \right)}{x'}\right]^{ab},
\end{equation}
the notation
\begin{equation}\label{Eqn30}
  \Tr[f] = \int\!(\uD{}x)\,f^{aa}(x,x),
\end{equation}
and the fact that $f^{bca}A^{b}_{k}=\uI(A_{k})^{ca}$, to rewrite
the second factor within the square brackets in (\ref{Eqn28}) as
\begin{equation}\label{Eqn31}
  \Tr\left\{-\uI {A'}_{\!\!k}\,\partial^{k}\frac{1}{\left[
  \Vec{\partial}^{2}-\uI{}g_{0} {A'}_{\!\!l}\,\partial^{l}
  \right]}\right\}.
\end{equation}
An elementary integration over $g_{0}$ from $0$ to some $g_{0}$
value then gives the familiar FP factor for $\Vacuum$ in
(\ref{Eqn28})
\begin{equation}\label{Eqn32}
  \det\left[1-\uI{}g_{0}\frac{1}{\Vec{\partial}^{2}}
  {A'}_{\!\!k}\,\partial^{k}\right].
\end{equation}

\section{Gauge Invariance and Modification of the FP Factor}\label{Section3}

Now consider the modification of the Lagrangian density
$\mathscr{L}$ in (\ref{Eqn02}):
\begin{equation}\label{Eqn33}
  \mathscr{L} \longrightarrow \mathscr{L}+\lambda
  \overline{\psi}\psi{}G^{a}_{\mu\nu}G_{a}^{\mu\nu}
  = \mathscr{L}_{1}
\end{equation}
which is obviously gauge invariant under the simultaneous local
gauge transformations in (\ref{Eqn07})--(\ref{Eqn09}).

The field equations corresponding to the Lagrangian density
$\mathscr{L}_{1\mathrm{T}}=\mathscr{L}_{1}+\mathscr{L}_{\mathrm{S}}$,
where $\mathscr{L}_{\mathrm{S}}$ is defined in (\ref{Eqn03}), are
given by
\begin{align}
  \nabla^{ab}_{\!\!\mu}\left(\left[1-4\lambda\overline{\psi}\psi
  \right]G_{b}^{\mu\nu}\Big.\right)
  &= -\left(\delta^{\nu}{}_{\sigma}\,\delta^{ac}-g^{\nu{}k}
  \partial_{k}D^{ab}\nabla^{bc}_{\!\!\sigma}\right)
  \nonumber \\
  &\qquad\quad{} \times\big[J_{c}^{\sigma}
  +g_{0}\overline{\psi}\gamma^{\sigma}t_{c}\psi\big]
  \label{Eqn34} \\[0.5\baselineskip]
  \bigg[\gamma^{\mu}\frac{\nabla_{\!\!\mu}}{\uI}
  -\lambda{}G^{a}_{\mu\nu}G_{a}^{\mu\nu}
  +m_{0}\bigg]\psi &= \eta
  \label{Eqn35} \\[0.5\baselineskip]
  \overline{\psi}\bigg[\gamma^{\mu}\frac{\overset{\lower0.7em
  \hbox{$\smash[t]{\scriptstyle\leftarrow}$}}{\nabla}_{\!\!\mu}
  \!{}^{*}}{\uI}+\lambda{}G^{a}_{\mu\nu}G_{a}^{\mu\nu}
  -m_{0}\bigg] &= -\overline{\eta}.
  \label{Eqn36}
\end{align}

The canonical conjugate momenta to $A_{a}^{1}$, $A_{a}^{2}$ are
given by
\begin{equation}\label{Eqn37}
  \pi_{a}^{i} = \left[1-4\lambda\overline{\psi}\psi\right]
  G_{a}^{i0}-\partial_{3}^{-1}\partial^{i}
  \left[1-4\lambda\overline{\psi}\psi\right]G_{a}^{30}
\end{equation}
$i=1,2$.   One may then express $G_{a}^{k0}$ as follows:
\begin{align}
  \left[1-4\lambda\overline{\psi}(x)\psi(x)\right]
  G_{a}^{k0}(x) &= \pi_{a}^{k}(x)-\partial_{k}\int\!(\uD{}x')\,
  D_{ab}(x,x')\Big[J_{b}^{0}(x')
  \nonumber \\
  &\qquad{} +g_{0}\overline{\psi}(x')\gamma^{0}t_{b}
  \psi(x')+{\nabla'}^{bc}_{\!\!\!\!j}\pi_{c}^{j}(x')\Big]
  \label{Eqn38}
\end{align}
$k=1,2,3$, with $\pi_{a}^{3}$ set equal to zero.

With the (independent) fields and their canonical conjugate
momenta kept fixed, we then have
\begin{equation}\label{Eqn39}
  \left[1-4\lambda\overline{\psi}(x)\psi(x)\right]
  \frac{\delta}{\delta{}J_{b}^{\nu}(x')}G_{a}^{k0}(x)
  = -\partial_{k}D_{ab}(x,x')\,\delta^{0}{}_{\nu}.
\end{equation}

The equal time commutation relations of the independent fields
$A_{a}^{1}(x)$, $A_{a}^{2}(x)$ are given by
\begin{equation}\label{Eqn40}
  \delta(x^{0}-{x'}^{0})\commute{A_{a}^{i}(x)}{\pi_{b}^{j}(x')}
  = \uI\delta_{ab}\delta^{ij}\delta^{4}(x-x')
\end{equation}
with $i,j=1,2$.   From the gauge constraint, we may then write
\begin{equation}\label{Eqn41}
  \delta(x^{0}-{x'}^{0})\commute{A_{a}^{k}(x)}{\pi_{b}^{l}(x')}
  = \uI\delta_{ab}\left[\delta^{kl}-\delta^{k3}
  \partial_{3}^{-1}\partial^{l}\right]\delta^{4}(x-x')
\end{equation}
with now $k,l=1,2,3$.

From (\ref{Eqn38}), (\ref{Eqn41}), we then obtain the commutation
relation
\begin{align}
  &\left[1-4\lambda\overline{\psi}(x)\psi(x)\right]
  \commute{A_{ka}(x')}{G_{a}^{k0}(x)}\delta(x^{0}-{x'}^{0})
  \nonumber \\
  &\qquad{} = 2\uI\delta_{aa}\delta^{4}(x-x')
  \nonumber \\
  &\qquad\quad{} -\partial_{k}\int\!(\uD{}x'')\,
  D_{ab}(x,x''){\nabla''}^{bc}_{\!\!\!\!\!\!j}
  \commute{A_{ka}(x')}{\pi_{c}^{j}(x'')}\delta(x^{0}-{x'}^{0}),
  \label{Eqn42}
\end{align}
where we recall that $D_{ab}(x,x'')$ involves the factor
$\delta(x^{0}-{x''}^{0})$.   The latter then implies that the last
term is given by
\begin{equation}\label{Eqn43}
  -\uI\partial_{k}\int\!(\uD{}x'')\,D_{ab}(x,x'')
  {\nabla''}^{ba}_{\!\!\!\!\!\!j}\left[\delta^{kj}
  -\delta^{k3}{\partial'_{3}}^{-1} {\partial'}^{j}\right]
  \delta^{3}(\Vec{x}'-\Vec{x}'')\,\delta(x^{0}-{x'}^{0}).
\end{equation}
Now we take the limit $\Vec{x}'\to\Vec{x}$ in the latter and
integrate over $\uD^{3}\Vec{x}$ to obtain
\begin{equation}\label{Eqn44}
  -\uI\int\!(\uD{}x'')\!\int\!\!\uD^{3}\Vec{x}\,
  \left[\partial_{j}-\partial_{j}\right]D_{ab}(x,x'')
  {\nabla''}^{ba}_{\!\!\!\!\!\!j}
  \delta^{3}(\Vec{x}-\Vec{x}'')\,\delta(x^{0}-{x'}^{0})
  = 0.
\end{equation}
This result will be used later in deriving the modification of the
FP factor.

The action principle gives
\begin{equation}\label{Eqn45}
  \frac{\partial}{\partial\lambda}\Vacuum = \uI\int\!(\uD{}x)
  \BK{0_{+}}{\overline{\psi}(x)\psi(x)G^{a}_{\mu\nu}(x)
  G_{a}^{\mu\nu}(x)}{0_{-}}.
\end{equation}
Consider the matrix element
\begin{align}
  \BK{0_{+}}{\big(G^{a}_{\mu\nu}(x)G_{a}^{\mu\nu}(x')\big)_{+}}{0_{-}}
  &= 2\BK{0_{+}}{\big(G^{a}_{k0}(x)G_{a}^{k0}(x')\big)_{+}}{0_{-}}
  \nonumber \\
  &\qquad{}
  +\BK{0_{+}}{\big(G^{a}_{kl}(x)G_{a}^{kl}(x')\big)_{+}}{0_{-}}.
  \label{Eqn46}
\end{align}
The second term is simply equal to
\begin{equation}\label{Eqn47}
  {G'}^{a}_{\!\!kl}(x){G'}_{\!\!a}^{kl}(x')\Vacuum
\end{equation}
expressed in terms of functional derivatives using our notation
below Eq.~(\ref{Eqn26}).   While to determine the first term, we
rewrite
\begin{equation}\label{Eqn48}
  G^{a}_{k0}(x) = \int\!(\uD{}z)\,\delta^{4}(x-z)
  \nabla^{ac}_{\!\!k}(z)A^{c}_{0}(z)-\int\!(\uD{}z)\,
  \delta^{4}(x-z)\,\partial^{z}_{0}A^{a}_{k}(z).
\end{equation}
We then have
\begin{align}
  &\BK{0_{+}}{\big(G^{a}_{k0}(x)G_{a}^{k0}(x')\big)_{+}}{0_{-}}
  \nonumber \\
  &\qquad{} = {G'}^{a}_{\!\!k0}(x){G'}_{\!\!a}^{k0}(x')\Vacuum
  \nonumber \\
  &\qquad\quad{}
  +\int\!(\uD{}z)\,\delta^{4}(x-z)\,\delta(z^{0}-{x'}^{0})
  \BK{0_{+}}{\commute{A^{a}_{k}(z)}{G_{a}^{k0}(x')}}{0_{-}}
  \nonumber \\
  &\qquad\quad{}
  -\uI\int\!(\uD{}z)\,\delta^{4}(x-z){\nabla'}^{ac}_{\!\!\!\!k}(z)
  \BK{0_{+}}{\frac{\delta}{\delta{}J_{c}^{0}(z)}G_{a}^{k0}(x')}{0_{-}}
  \label{Eqn49}
\end{align}
where the second term comes from the non-commutativity of the time
derivative and the time ordering operation as resulting from the
last term in (\ref{Eqn48}), and the third term follows from the
rule of functional differentiation in (\ref{Eqn27}) as resulting
from the first integral in (\ref{Eqn48}).

From (\ref{Eqn38}), (\ref{Eqn42}), (\ref{Eqn44}), the right-hand
side of (\ref{Eqn49}) simplifies for $x'\to{}x$ to
\begin{equation}\label{Eqn50}
  \left[{G'}^{a}_{\!\!k0}(x){G'}_{\!\!a}^{k0}(x)
  +\varDelta'(x)\right]\Vacuum
\end{equation}
where
\begin{align}
  \varDelta'(x) &= 2\int\!(\uD{}z)\,\frac{\delta^{4}(z-x)}{\left[
  1-4\lambda\overline{\psi}'(x)\psi'(x)\right]}K'(x,z)
  \label{Eqn51} \\
  K'(x,z) &= \uI\left[\delta_{aa}\delta^{4}(0)+\frac{1}{2}
  \partial^{x}_{k}{\nabla'}^{ac}_{\!\!\!\!k}(z)D'_{ac}(x,z)\right]
  \label{Eqn52}
\end{align}
involving a familiar $\delta^{4}(0)$ term.

All told, the expression (\ref{Eqn45}) becomes
\begin{align}
  \frac{\partial}{\partial\lambda}\Vacuum &= \uI\int\!(\uD{}x)\,
  \overline{\psi}'(x)\psi'(x){G'}^{a}_{\!\!\mu\nu}(x)
  {G'}_{\!\!a}^{\mu\nu}(x)\Vacuum
  \nonumber \\
  &\qquad{}
  +2\uI\int\!(\uD{}x)\,
  \overline{\psi}'(x)\psi'(x)\varDelta'(x)\Vacuum
  \label{Eqn53}
\end{align}
which upon an elementary integration over $\lambda$ leads to
\begin{equation}\label{Eqn54}
  \Vacuum = \uE^{\uI{}M'}\exp\left[\uI\lambda\!\int\!(\uD{}x)\,
  \overline{\psi}'(x)\psi'(x){G'}^{a}_{\!\!\mu\nu}(x)
  {G'}_{\!\!a}^{\mu\nu}(x)\right]\Vacuum_{\lambda=0}
\end{equation}
where
\begin{equation}\label{Eqn55}
  M' = -\int\!(\uD{}z)\,\delta^{4}(x-z)\,\ln\left[1
  -4\lambda\overline{\psi}'(x)\psi'(x)\right]K'(x,z)
\end{equation}
and $\Vacuum_{\lambda=0}$ is the vacuum-to-vacuum amplitude
corresponding to the Lagrangian density $\mathscr{L}_{\mathrm{T}}$
in (\ref{Eqn01}) involving the FP factor in (\ref{Eqn32}).   That
is, the familiar FP factor gets modified by a multiplicative
factor $\exp[\uI{}M']$ for the gauge invariant Lagrangian density
$\mathscr{L}_{1}$ in (\ref{Eqn33}).

\section{Gauge Breaking Interactions}\label{Section4}

In the present section we consider the addition of a gauge
breaking term to the Lagrangian density $\mathscr{L}$ in
(\ref{Eqn02}).   It is well known if the addition of the simple
source term $\mathscr{L}_{\mathrm{S}}$ in (\ref{Eqn03}) to
$\mathscr{L}$ cause difficulties (cf. Rivers~\cite{Rivers_1987},
p.~204) in the quantization problem in the path integral formalism
as the action $\int(\uD{}x)\mathscr{L}_{\mathrm{T}}(x)$, with
$\mathscr{L}_{\mathrm{T}}(x)$ in (\ref{Eqn01}), is not gauge
invariant.   We will see how easy it is to handle the addition of
a gauge breaking term even to $\mathscr{L}_{\mathrm{T}}$.

Consider the Lagrangian density
\begin{equation}\label{Eqn56}
  \mathscr{L}_{2\mathrm{T}} = \mathscr{L}_{\mathrm{T}}
  +\frac{\lambda}{2}A^{a}_{\mu}A_{a}^{\mu}\overline{\psi}\psi.
\end{equation}
Then an analysis similar to the one in Sect.~\ref{Section3} shows
that
\begin{equation}\label{Eqn57}
  G_{a}^{k0} = \pi_{a}^{k}-\partial_{k}D_{ab}\left[J_{b}^{0}
  +\lambda{}A_{b}^{0}\overline{\psi}\psi+g_{0}\overline{\psi}
  \gamma^{0}t_{b}\psi+\nabla^{bc}_{\!\!\nu}\pi_{c}^{\nu}\right].
\end{equation}
Using the fact that
\begin{equation}\label{Eqn58}
  \partial_{k}G_{a}^{k0} = \nabla^{ab}_{\!\!k}\partial_{k}A_{b}^{0}
\end{equation}
we obtain upon multiplying (\ref{Eqn57}) by
\begin{equation*}
  \nabla^{ca}_{\!\!l}\partial^{l}\frac{1}{\Vec{\partial}^{2}}
  \partial_{k}
\end{equation*}
and using (\ref{Eqn14}), we obtain
\begin{equation}\label{Eqn59}
  \left(\nabla^{ca}_{\!\!l}\partial^{l}\frac{1}{\Vec{\partial}^{2}}
  \nabla^{ab}_{\!\!k}\partial_{k}\right)A_{b}^{0}
  = -J_{c}^{0}-\lambda{}A_{c}^{0}\overline{\psi}\psi+\ldots
\end{equation}
where the dots correspond to terms \emph{independent} of
$J_{b}^{0}$ and $A_{b}^{0}$. We introduce the Green operator
$N^{be}(x,x')$ satisfying
\begin{equation}\label{Eqn60}
  \left[\nabla^{ca}_{\!\!l}\partial^{l}\frac{1}{\Vec{\partial}^{2}}
  \nabla^{ab}_{\!\!k}\partial_{k}+\lambda\delta^{cb}
  \overline{\psi}(x)\psi(x)\right]N^{be}(x,x') =
  \delta^{ce}\delta^{4}(x-x')
\end{equation}
to obtain from (\ref{Eqn59})
\begin{equation}\label{Eqn61}
  \frac{\delta}{\delta{}J_{b}^{0}(x)}A_{b}^{0}(x)
  = -N^{bb}(x,x).
\end{equation}

Hence the action principle together with (\ref{Eqn61}) gives
\begin{align}
  \frac{\partial}{\partial\lambda}\Vacuum &= \frac{\uI}{2}
  \int\!(\uD{}x)\,{A'}^{a}_{\!\!\mu}(x){A'}_{\!\!a}^{\mu}(x)
  \overline{\psi}'(x)\psi'(x)\Vacuum
  \nonumber \\
  &\qquad{}
  -\frac{1}{2}\int\!(\uD{}x)\,
  \overline{\psi}'(x)\psi'(x){N'}^{bb}(x,x)\Vacuum.
  \label{Eqn62}
\end{align}

Upon integrating the latter over $\lambda$, by using in the
process (\ref{Eqn60}), we obtain
\begin{align}
  \Vacuum &= \exp\left[-\frac{1}{2}\Tr\ln\left(1
  +\frac{\lambda}{\nabla'_{\!\!l}\partial^{l}
  (\Vec{\partial}^{2})^{-1}\,\nabla'_{\!\!k}
  \partial_{k}}\overline{\psi}'\psi'\right)\right]
  \nonumber \\
  &\qquad{}
  \times\exp\left[\uI\frac{\lambda}{2}\int\!(\uD{}x)\,
  {A'}^{a}_{\!\!\mu}(x){A'}_{\!\!a}^{\mu}(x)
  \overline{\psi}'(x)\psi'(x)\right]\Vacuum_{\lambda=0}
  \label{Eqn63}
\end{align}
with an obvious modification of the FP factor with the latter
occurring in $\Vacuum_{\lambda=0}$.

\section{Conclusion}\label{Section5}

The quantum action (dynamical) principle leads systematically to
the FP of non-abelian gauge theories with no much effort.   It is
emphasized, in the process of the analysis, that no restrictions
may be set on the external current $J^{a}_{\mu}$, coupled to the
gauge field $A_{a}^{\mu}$ (such as
$\partial^{\mu}\!J^{a}_{\mu}=0$), until all functional
differentiations with respect to it are taken so that all of its
components may be varied independently.   We have considered gauge
invariant as well as gauge non-invariant interactions and have
shown that the FP factor needs to be modified in more general
cases and expressions for these modifications were derived.   [It
is well known that even the simple gauge breaking source term
$\mathscr{L}_{\mathrm{S}}$ in (\ref{Eqn03}) causes complications
in the path integral formalism. The path integral may, of course,
be readily derived from the action principle.]    The presence of
the source term $\mathscr{L}_{\mathrm{S}}$ in the Lagrangian
density is essential in order to generate the Green functions of
the theory from the vacuum-to-vacuum transition amplitude, as a
generating functional, by functional differentiations.   We have
also shown, in particular, that a gauge invariant theory does not
necessarily imply the familiar FP factor for proper quantization.
Finally we note that even abelian gauge theories, as obtained from
the bulk of the paper by taking the limit of $f^{abc}$ to zero and
replacing $t^{a}$ by the identity, may lead to modifications, as
multiplicative factors in $\Vacuum$, as clearly seen from the
expressions in (\ref{Eqn55}) and (\ref{Eqn63}).

\section*{Acknowledgment}

The authors would like to acknowledge with thanks for being
granted t\href{http://rgj.trf.or.th/}{the ``Royal Golden Jubilee
Ph.D. Program''} by \href{http://www.trf.or.th/}{the Thailand
Research Fund} (Grant No. PHD/0117/2545) for partly
carrying out this project.

\end{document}